\documentstyle[12pt,epsfig]{article}
 \newcommand{\p}{{\bf p}}
 \newcommand{\q}{{\bf q}}
 \newcommand{\so}{\stackrel{o}{\bf s}}
 
 \newcommand{\bK}{{\bf K}}
 \newcommand{\0}{{\bf 0}}
 
 \newcommand{\J}{{\bf J}}
 \newcommand{\lam}{{\lambda}}
 \newcommand{\ra}{{\rangle}}
 \newcommand{\la}{{\langle}}
 \newcommand{\be}{\begin{equation}}
 \newcommand{\ben}{\begin{eqnarray}}
 \newcommand{\een}{\end{eqnarray}}
 \newcommand{\ee}{\end{equation}}
 
 \newcommand{\po}{\stackrel{o}{\p}}
 \newcommand{\dsty}{\displaystyle}

 \newcommand{\ponot}{\stackrel{\not\,{o}}{\p}}
 \newcommand{\ponotnot}{\stackrel{\not\,\not{o}}{\p}}
 \newcommand{\poc}{\stackrel{o}{ p}}
 \newcommand{\ponotc}{\stackrel{\not\,{o}}{p}}

 \newcommand{\pmo}{\stackrel{\not\,{o}}{\p\smash{\lefteqn{_1}}}\phantom{1}}
 \newcommand{\pno}{\stackrel{\not\,{o}}{\p\smash{\lefteqn{_2}}}\phantom{1}}

 \newcommand{\DDo}{D\smash{\lefteqn{^{(1/2)}_{\lam_{s_1}\lam'_{s_1}}}}\phantom{aaaa}}
 \newcommand{\DDd}{D\smash{\lefteqn{^{(1/2)}_{\lam_{s_2}\lam'_{s_2}}}}\phantom{aaaa}}

\newcommand{\Dp}{\stackrel{+}{D\smash{\lefteqn{^{(1/2)}}}}\phantom{aa1}}
\newcommand{\D}{D\smash{\lefteqn{^{(1/2)}}}\phantom{aa1}}
\newcommand{\Dtss}{D\smash{\lefteqn{^{(1/2)}_{\lam'_{s_3}\lam''_{s_3}}}}\phantom{aaa1}}

\begin{document}

\begin{center}
{\bf
Nucleon Magnetic Moments in the Quasipotential Quark Model
}
\\
 T.P.Ilichova, S.G.Shulga
\\[6pt] {\it Francisk
Skaryna Gomel State University, LPP JINR, Dubna} \\
e-mail: shulga@cv.jinr.ru, ilicheva@cv.jinr.ru
\end{center}

\begin{abstract}
\par
  Calculation of the nucleon magnetic moments was performed 
in the quasipotential quark model. 
  The main contribution into the nucleon magnetic moments is 
determined by the $SU(6)$ symmetrical spin-isospin part of 
the wave function and is equal to 3 and -2 (in the units 
of nuclear magnetons) for proton and neutron, respectively.
  Corrections are explained by relative quark motion and are
equal to $-4a/3$ and $+a$ for proton and neutron, respectively, 
where the dimensionless value $a$ depends on the dimensionless 
ratio of the oscillator parameter and quark mass in case of 
the oscillator interaction.
  It is shown that the independence of the nucleon magnetic 
moments on nucleon mass is a consequence of relativistic 
kinematics.
 The invariant dimensionless parameter is introduced to estimate
the violation of current conservation.
 By means of this parameter it is shown that the current leakage
is neglegibly small in the model.

\end{abstract}

\begin{center}
{\bf Introduction.}
\end{center}

\par
  This paper is dedicated to calculation of nucleon
magnetic moments in the relativistic quasipotential
quark model~\cite{IzvVuzov_Ilicheva}.
  In the nonrelativistic quark model the nucleon magnetic moments 
are given by:
$\mu^p=M/m,\;\mu^n=(-2/3)M/m$
(in units of the nuclear magnetons $e/2M$,
$e$ electron charge, $M$ nucleon mass).
  To obtain the experimental values of the nucleon magnetic moments
it is needed to introduce the relation of nucleon and quark mass
$M\approx 3m$.
  In ~\cite{BogolyubovPN} the expresions
$\mu^p=(1-b)M/E_0,\;\mu^n=(-2/3)(1-b)M/E_0$,
containing the ratio of the nucleon quark mass and 
level of quark energy $E_0$,
are obtained on the base of the relativistic shell model
of quasiindependent quarks being in the potential hole.
  The model of quasiindependent quarks solves the 
confinement problem by means of a big quark mass and big 
bound energy.
  But in this case a quark has a big anamalous magnetic moment
($M/m-M/E_0$) and relation with deep inelastic processes is lost.
  In the bag quark model~\cite{close} it is shown that dirac particle, 
closed in the bag with radius $R$, has not such big anamalous 
magnetic moments because of the big quark energy.
  In this model the quark and proton magnetic moments
are proportional to the bag radius ($\mu^p\approx 0.4RM$) and
the quark mass may be small.
  The relativistic approach in paper~\cite{Gerasimov82}  proposes
the assumption of stationarity of the one particle energy in 
a system of interactive quarks. 
  In this case the nucleon magnetic moments depend 
on "effective qaurk mass" $\varepsilon_q$.
  In light-front quark model~\cite{Berest_Terent76,Aznauryan80} 
the nucleon magnetic moments are proportional to ratio $M/m$.
\par
  The relative quark motion in nucleon is defined by at least two parameters:
constituent quark parameter ($m,E_0,\varepsilon_q$) and
energy scale parameter $\gamma$ ($\gamma=1/R$ in case of the quark 
bag model).
  The nucleon mass is an additional parameter. 
    In all models considered above, as in other relativistic
models
(see for example~\cite{Gerasimov82,Berest_Terent76,Aznauryan80,chung_coester,coester_dannbom_riska}),
the dimensionless nucleon magnetic moments (in units of the
nuclear magnetons) are
defined by two dimensionless parameters:
combination of the model relative motion parameters ($m/\gamma$) and
ratio of the one relative motion parameter
and the nucleon mass ($M/m, M/E_0,M/\varepsilon_q$).
  This work  will  show that in the quasipotential model
the nucleon magnetic moments don't depend on the nucleon mass.

\newpage
\begin{center}
{\bf 1. Basic relations of the model.}
\end{center}
\par
  In work~\cite{IzvVuzov_Ilicheva} a structure of the three quark nucleon
wave function is considered in the framework of the
quasipotential approach.
In particular the relation of the three quark component of the nucleon
vector state decomposition in the Fock-momentum space was obtained
(summary of notations will be done in the end of this section):
\be
\la
\{\p_k \lam_{s_k} \lam_{t_k}\}
\mid K \lam_J \lam_T\ra
=
(2\pi)^3
\delta^{(3)}(\sum_{k=1}^3\po_k)
\left[
\prod_{k=1}^{3}
\stackrel{+}{D}\smash{^{(1/2)}_{\lam'_{s_k}\lam_{s_k}}}
(L^{-1}_K,p_k)
\right]
\Phi^{(+)}_{M\lam_J\lam_T\{\lam'_{s_k}\lam_{t_k}\}}(\{\po_k\}).
\label{phiplus_chrz_comp}
\ee
If we perform summation of the quark spin by scheme [[1+2]+3]
and assume that the orbital angular momentum of the
system [1+2] is equal to zero
in [1+2] center-of-mass system (c.m.s.) and
the orbital angular momentum of the
third quark relatively the [1+2] system
in the nucleon c.m.s. is equal to zero, 
then function $\Phi^{(+)}$ has the following form:
\ben
\Phi^{(+)}_{\{\lam_{s_k}\lam_{t_k}\} }
(\{\po_k\})  =
(2\pi)^3
\left[
\DDo(L_{\po_{12}},\pmo)
\DDd(L_{\po_{12}},\pno)
\delta_{\lam_{s_3}\lam'_{s_3}}
\right]
\chi_{\{\lam'_{s_k}\lam_{t_k}\}}
\varphi_M(\{\po_k\}),
\label{Shirokov91}
\een
where
$
\chi_{\{\lam_{s_k}\lam_{t_k}\}}=
\left[
\xi^{MS}_{\{\lam_{s_k}\}}
\eta^{MS}_{\{\lam_{t_k}\}}
+
\xi^{MA}_{\{\lam_{s_k}\}}
\eta^{MA}_{\{\lam_{t_k}\}}
\right]/\sqrt{2}
$
- $SU(6)$
nucleon wave function is symmetrical relatively the quark permutations [3].
We will omit indices $\lam_J$ and $\lam_T$ anywhere supposing
$\lam_J=+1/2$, and  we will indicate
the proton as index $p$ ($\lam_T=+1/2$) or the neutron as index $n$
($\lam_T=-1/2$). Three Wigner rotation matrices $D$
in~(\ref{phiplus_chrz_comp}) are related with transition from the
laboratory system to c.m.s. of nucleon. Two Wigner rotation
matrices $D$ in expression~(\ref{Shirokov91}) are related
with transition from the rest frame system [1+2] to c.m.s. of the the
nucleon.
Function $\varphi_M$ is normalized according to the relation:
\ben
\int
d\Omega_{\po_1}d\Omega_{\po_2}
\mid
\varphi_M(M_0(\{\po_k\}))
\mid^2
/E_{\po_3}=1.
\label{norm_varphi}
\een
\par
In work~\cite{IzvVuzov_Ilicheva} 
the relativistic three-particle oscillator was proposed
on the basis of the quasipotential equation and its approximate
solution was obtained:
\be
\varphi_M^{osc}(\{\po_k\})
\approx N\exp[-m(\sum_{k=1}^3E_{\po_k}-3m)/\gamma^2].
\label{varphi_approx_eff_m}
\ee
If the effective mass approximation is better then
the approximation in~(\ref{varphi_approx_eff_m}) is also better.
The effective mass approximation supposes that
$\delta_E^{(k)}\equiv m^{eff}_k-\la E_{p_k}\ra$ is small,
where $m^{eff}_k=\sqrt{m^2+\la p^2_k\ra}$,
$\;\la p^2_k\ra=\int
d\Omega_{\p_1}d\Omega_{\p_2}
|
\varphi_M
|^2
p^2_k
/E_{\po_3}$,
$\;\la E_{p_k}\ra=\int
d\Omega_{\p_1}d\Omega_{\p_2}
|
\varphi_M
|^2
E_{p_k}
/E_{\po_3}
$.
We suppose that quarks have equal masses.
Symmetry relatively the quark permutations
in normalization condition~(\ref{norm_varphi}) leads to the following:
$m^{eff}_k,\,\la E_{p_k}\ra,\,\delta_E^{(k)}$ don't depend on
quark number $k$. This symmetry relatively the quark permutations will be
important to obtain the main exppression of this paper
(\ref{mu_approx})-(\ref{a_approx}).
\par
Here we give a list of notations:
$d\Omega_{\p_k}=d\p_k/E_{\p_k}$,
$E_{\p_k}=\sqrt{\p_k^2 + m^2}$,
$E_K=\sqrt{\bK^2+M^2}$,
$\bK$ and $\p_k$ are nucleon and quark momenta,
$M$ and $m$ are nucleon and quark masses,
$U$ and $u$ are nucleon and quark spinors,
$\{\p_k \lam_{s_k}\lam_{t_k}\}\equiv
\p_1\lam_{s_1}\lam_{t_1}
\p_2\lam_{s_2}\lam_{t_2}
\p_3\lam_{s_3}\lam_{t_3}.$
The total spins and isospins of the nucleon and constituent quarks
are not indicated and equal to $1/2$;
$\lam_J$, $\lam_{s_k}$ are the third projections
of the nucleon and quark spins.
We assume a sum over the repeated indices of
the third projection of spin
and isospin;
$\poc_k=(L^{-1}_K p_k)$,
$\ponotc_k=(L^{-1}_{\ponotc_{12}}\poc_k)$,
$\poc_{12} = \poc_1 + \poc_2$,
$L^{-1}_K$ and $L^{-1}_{\ponotc_{12}}$ are the Lorentz boosts to the
rest frame of nucleon and [1+2] quarks.
The normalization condition of vector states and
spinors of nucleon and quarks are given by:
$$
\la K' J' \lam'_J T' \lam'_T
\mid K J \lam_J T \lam_T\ra
=
(2\pi)^3
\delta^3(\bK'-\bK)
\delta_{\lam'_J \lam_J}
\delta_{\lam'_T \lam_T},
$$
$$
\la
\p'_k s_k \lam'_{s_k} t_k \lam'_{t_k}
\mid
\p_k s_k \lam_{s_k} t_k \lam_{t_k}
\ra
=
(2\pi)^3
E_{\p_k}
\delta^3(\p'_k-\p_k)
\delta_{\lam'_{s_k} \lam_{s_k}}
\delta_{\lam'_{t_k} \lam_{t_k}},
$$
\be
\bar{u}^{\lam_{s_k}}(\p)
u^{\lam_{s_k}}(\p)
=
m,
\;\;
\bar{U}^{\lam_J}(\bK)
U^{\lam_J}(\bK)
=
M/E^M_{\bK}
\mbox{ (there is no sum over } \lam_{s_k} \mbox{ and } \lam_J).
\label{normir}
\ee

\begin{center}
{\bf 2. Nucleon electromagnetic current}
\end{center}
\par
 We suppose that in the three quark model the nucleon current
is expressed as a sum of quark currents
$\hat{J}(x) = \sum_{k=1}^3\hat{j}^{(k)}(x)$,
and write the current matrix element between nucleon states
in the rest frame of the initial nucleon:
$$
J_\mu^{\lam_J\lam'_J}\equiv
\la E^M_{\bK} \bK \lam'_J
\lam'_T\mid J(0)\mid M\0 \lam_J\lam_T\ra =
\frac{1}{(2\pi)^{18}}
\int\left[\prod^3_{k=1}d\Omega_{\p_k} d\Omega_{\p'_k}\right]
\times
$$
\be
\times
\la E^M_{\bK}\bK \lam'_J\lam'_T
|\{\p'_k\lam'_{s_k}\lam'_{t_k}\} \ra
\la\{\p'_k\lam'_{s_k}\lam'_{t_k}\}|
\sum_{k=1}^3 j^{(k)}(0)
|\{\p_k\lam_{s_k}\lam_{t_k}\}\ra
\la\{\p_k\lam_{s_k}\lam_{t_k}\}
| M\0 \lam_J\lam_T\ra.
\label{matr_elem_1}
\ee
Since the three quark component of the nucleon vector state
is symmetrical relatively the quark permutations, 
it is possible to replace
$\sum_{k=1}^3\hat{j}^{(k)}(0)$ to
$3\hat{j}^{(3)}(0)$ and it is possible to use
the scheme of decomposition of the wave function (WF)
over states with the definite angular momentum,
assuming $\p_1$ and $\p_2$
as independent variables.
Involving WF~(\ref{phiplus_chrz_comp})
in~(\ref{matr_elem_1}), we obtain:
\be
 J_{\mu}(\bK,{\0}) =
 3 \int d\Omega_{\p_1} d\Omega_{\p_2} \varphi(\po_1,\po_2,\po'_3)
(\chi M^{(1)} M^{(2)} \Gamma^{(3)}_{\mu} \chi)
\varphi(\p_1,\p_2,\p_3)/E_{\p_3}E_{\po'_3} ,
\label{current}
\ee
where indices $\lam_J=\lam'_J=+1/2$ are omitted and
the following matrix notations are used:
\ben
&& M^{(k)}_{\lam'_{s_k}\lam'_{t_k}\lam_{s_k}\lam_{t_k}} \equiv
\delta_{\lam'_{t_k}\lam_{t_k}}
 \Bigl[
\Dp(L_{\po_{12}},\ponot_k)
\D(L^{-1}_\bK,\p_k)
\D(L_{\p_{12}},\ponotnot_k)
\Bigr]_{\lam'_{s_k}\lam_{s_k}}, \quad k = 1,2;
\nonumber\\
&& \Gamma^{(3)}_{_{\mu},\lam'_{s_3}\lam'_{t_k}\lam_{s_3}\lam_{t_3}}
\equiv
\Dtss(L^{-1}_\bK,\p'_3)
j_{\mu}^{(3)\lam''_{s_3}\,\lam_{s_3}}(\p'_3 ,\p_3)
\delta_{\lam'_{t_3}\lam_{t_3}}
\nonumber\\
&&  j_{\mu}^{(3)\,\lam'_{s_3}\lam_{s_3}}(\p'_3,\p_3)
\delta_{\lam'_{t_3}\lam_{t_3}} \equiv
\la\p'_3\lam'_{s_3}\lam'_{t_3}|j^{(3)}_{\mu}(0)|\p_3\lam_{s_3}\lam_{t_3}
\ra.
\nonumber
\een
The current normalization condition
$J_0(\0,\0)=e_N$
($e_N$ is the nucleon charge in units of the electron charge $e$)
leads to the WF normalization
condition~(\ref{norm_varphi}) ~\cite{IzvVuzov_Ilicheva}, which was obtained by
the Green function method.
Quark momenta are related by:
$\p_3=-\p_1-\p_2$,
$\po'_3=-\po'_1-\po'_2$,
$\p'_1=\p_1$,
$\p'_2=\p_2$,
$\ponotnot_k=L^{-1}_{p_{12}}\p_k$,
$\p_{12}=\p_1+\p_2.$
\par
Let us consider the nucleon current conservation law.
If $\so=0$, then
$s=\stackrel{o}{s}_0 K/M$ for arbitrary 4-momentum $s$.
Assuming $K=(E_K,\bK)$, $K'=(M,\0)$, we have
$$
p_1+p_2+p'_3=K(E_{\po_1}+E_{\po_2}+E_{\po'_3})/M,
\quad
p_1+p_2+p_3=K'(E_{\p_1}+E_{\p_2}+E_{\p_3})/M.
$$
For transfer momentum $q=K'-K$, we obtain
\be
(E_{\po_1}+E_{\po_2}+E_{\po'_3})q_\mu/M=
(p'_3-p_3)_\mu-(E_{\po'_3}-E_{\p_3})(1,\0)_\mu.
\label{transf_mom_rel}
\ee
If quark current is conserved, then the last term
in~(\ref{transf_mom_rel}) represents a deviation from the current
conservation law in the model.
To estimate the current conservation numerically, we introduce
the dimensionless invariant ratio
\be
\delta_J=
\left|
\frac{q_0 J_0-\q\J}{q_0 J_0+\q\J}
\right|
\times 100 \%.
\label{delta_J}
\ee
According to~(\ref{transf_mom_rel}) the current~(\ref{current})
is conserved
better if the effective mass approximation is satisfied better
(see p.1 in~\cite{IzvVuzov_Ilicheva}), where quark energies
$E_{\p_3}$ and $E_{\po'_3}$
of the initial and final nucleons in the nucleon c.m.s.
are replaced by effective quark mass $m^{eff}$.
The value $\delta_J$ allows to estimate
a part of the current lost in the model
with fixed number of particles.
It was obtained that $\delta_J<0.1\%$ for
$t\in[0.0001,2]$ $GeV^2$  ($t=|q^2|$) for
ratio $\gamma/m=0.55$, which corresponds to the best fit of the nucleon
magnetic moments.
Thus, the current leakage which is related to the assumption of the
fixed number of particles is neglegibly small.
\par
The Dirac form factors of the nucleon are given by the expression:
\be
 J_{\mu}^{\lam'_J\lam_J}(\bK,\0) =
 {\dsty e_N} \bar U^{\lam'_J}(\bK)
 \{ \gamma_{\mu}F_1(t)
 + \frac{i\kappa}{2M} F_2(t)
 \sigma_{\mu\nu} q_{\nu}
         \}
 U^{\lam_J}({\0}),
 \label{ffactor def}
\ee
where $\kappa$ is the nucleon anomalous magnetic moment,
$
\sigma_{\mu\nu} = \frac{i}{2}(\gamma_{\mu}\gamma_{\nu} -
\gamma_{\nu}\gamma_{\mu})
$.
For convenience we introduce the polarization 3-vectors
$\varepsilon_{(k)}$, which have components
$\varepsilon_{(k)i} = \delta_{ki}.$
For the 3-current we have
\ben
 \left( \varepsilon_{(k)}\, {\bf J}^{\lam'_J\lam_J}
 (\bK,{\0}) \right)=
  \frac{\dsty e_N}
  {\dsty \sqrt{2E^M_{\bK}(M+ E^M_{\bK})}}
\left[
(\bK \varepsilon_{(k)}) G_E(t)
 + i({\bf\sigma}[\bK\times \varepsilon_{(k)}]) G_M(t)
 \right]_{\lam'_J\lam_J},
\label{ek*J_GEGM}
\een
where the Sacks form factors have the form:
$G_E(t) = F_1(t) + F_2(t) t/4M^2$,
$G_M(t)= F_1(t) + \kappa F_2(t)$.
It has been numerically shown, that the imaginary part of form factors,
defined by~(\ref{ek*J_GEGM}) for current~(\ref{current}) is
neglegibly small. So, the constructed current approximately satisfies
 $T$-invariance with the sufficient precision.
\begin{center}
{\bf 3. Nucleon magnetic moments.}
\end{center}
\par
To write the analitical expression for the magnetic moment,
we differentiate~(\ref{ek*J_GEGM}) over $\bK_1$,
assuming $\bK=0$, $\lam'_J=\lam_J=1/2$; $k=2$, $G_M(0)=\mu_N$
and using~(\ref{current}):
\be
 \mu_N \frac{\dsty e_N}{\dsty 2 M}\,
 = -i\frac{d}{d\bK_1}I|_{\bK=0},
\label{mag_mom}
\ee
$$
I= 3 \int d\Omega_{\p_1} d\Omega_{\p_2} \varphi(\po_1,\po_2,\po'_3)
(\chi M^{(1)} M^{(2)} \hat{e}_3G^{(3)}\chi)
\varphi(\p_1,\p_2,\p_3)/E_{\p_3}E_{\po'_3} ,
$$
$$
\hat{e}_3G^{(3)}
\equiv
\varepsilon_{(2)k}\Gamma^{(3)}_{k}=
\Dtss(L^{-1}_\bK,\p'_3)
\left(\varepsilon_{(2)k}
{\bf j}_{k}^{(3)}(\p'_3 ,\p_3)
\right)
$$
$$
(\chi M^{(1)} M^{(2)} \hat{e}_3G^{(3)}\chi)=
\frac{1}{2}
\left[
(\eta^{MS} \hat{e}_3 \eta^{MS})
(\xi^{MS} M^{(1)} M^{(2)} G^{(3)} \xi^{MS}) +
 \right.
$$
\be
 \left.
+ (\eta^{MA} \hat{e}_3 \eta^{MA})
(\xi^{MA} M^{(1)} M^{(2)} G^{(3)} \xi^{MA})
 \right].
\label{Matr-elem}
\ee
  Wave function $\varphi_M$ is 3-scalar 
and depends on $\p^2_K$ and $(\p_k\p_s)$
($k,s=1,2$ is the quark number).
  From such function we have
\be
\int d\Omega_{\p_1} d\Omega_{\p_2}
|\varphi(\{\p_k\})|^2
\p_k^i\p_s^j\sim\delta_{ij},\quad
\int d\Omega_{\p_1} d\Omega_{\p_2}
\varphi(\{\p_k\})
\p_k^i
\nabla_{\p_s^j}\varphi(\{\p_k\})
\sim\delta_{ij}.
\label{condition_varphi_delta}
\ee
Taking into account~(\ref{condition_varphi_delta})
it is followed from~(\ref{mag_mom}) that
\be
 \mu_N \frac{\dsty e_N}{\dsty 2 M}\,
=
3 \int d\Omega_{\p_1} d\Omega_{\p_2}
|\varphi(\{\p_k\})|^2 (-i)\frac{d}{d\bK_1}
(\chi M^{(1)} M^{(2)} \hat{e}_3G^{(3)}\chi)|_{\bK=0}.
\label{mag_mom_after_condition}
\ee
Dependence on $\bK$ in~(\ref{mag_mom_after_condition})
under the derivative sign
is contained in form of a complicated function
in the relative quark momenta in the nucleon c.m.s.,
in the Wigner rotation matrices
and in the quark current. Using the explicit form of $\po_k$,
$D$-matrices
and quark current, we will show below, that the right-handed side of
equation~(\ref{mag_mom_after_condition}) is proportional to the
first order
of ratio $1/M$. So, $\mu_N$ in~(\ref{mag_mom_after_condition})
does not depend on the nucleon mass,
and it is a consequence of relativistic kinematics.
\par
The quark current with charge operator $\hat{e}_3$ has the form:
$ j_{\mu}^{(3)}(\p'_3,\p_3) =
 \hat{e}_3  {\bar u}(\p'_3) \gamma_{\mu} u(\p_3) $.
Let us rewrite it in the 2-spinors:
\be
( \varepsilon_{(2)} {\bf j}^{(3)}(\p'_3,\p_3)) =
  \frac{\hat{e}_3}{2}
 \left[
  (\sigma^{(3)} \varepsilon_{(2)})(\sigma^{(3)} \p_3)
  \frac{\sqrt{(E_{\p'_3}+m)}}{\sqrt{(E_{\p_3}+m)}} +
  (\sigma^{(3)} \p'_3)(\sigma^{(3)} \varepsilon_{(2)})
  \frac{\sqrt{(E_{\p_3}+m)}}{\sqrt{(E_{\p'_3}+m)}} +
 \right].
 \label{q_curr}
\ee
The upper index of Pauli matrices $\sigma^{(k)}$ indicates the
quark number.
Using the explicit form of Lorentz transformation,
we obtain following relations in
first order of $\bK$:
$
  \p'_3 \approx \p_3 +
  (E_{\p_1} + E_{\p_2} + E_{\p_3})
   \bK/M,
  \;
  E_{\p'_3} \approx E_{\p_3} +
  (\bK \p_3)(E_{\p_1} + E_{\p_2} + E_{\p_3})/M E_{\p_3}.
$
As a result in the first order of $\bK$ assuming $\bK=(K,0,0)$,
we obtain the following from~(\ref{q_curr}):
\ben
( \varepsilon_{(2)} {\bf j}^{(3)}(\p'_3,\p_3)) =
 i\frac{\dsty \hat{e}_3 }{\dsty 2 M}
 (\sigma^{(3)}[\bK\times \varepsilon_{(2)}])
  (E_{\p_1} + E_{\p_2} + E_{\p_3})-
\nonumber\\
- i\frac{\hat{e}_3}{2 M}
(\sigma^{(3)}[\p_3\times \varepsilon_{(2)}])
\,(\bK \p_3)
 \frac{
  (E_{\p_1} + E_{\p_2} + E_{\p_3})
 }{ (E_{\p_3}+m) E_{\p_3}}
  + \hat{e}_3(\varepsilon_{(2)}\p_3) + O(\bK^2).
 \label{q_curr_apprx}
\een
From the terms of expression~(\ref{q_curr_apprx}) we have three
contributions into the nucleon magnetic moment:
\ben
  \mu_N = \mu_1 +  \mu_2 +  \mu_3.
\label{three_part_mag_mom}
\een
According to~(\ref{q_curr_apprx}) the first
term of~(\ref{three_part_mag_mom}) contains
$ (\sigma^{(3)}[\bK\times \varepsilon_{(2)}])
$ and corresponds to the quark motion as a 
part of nucleon with momentum $\bK$. 
The second term contains
$(\sigma^{(3)}[\p_3\times \varepsilon_{(2)}])$
and takes into account the relative quark motion with momentum $\p_3$.
The third term, as it is shown below,
takes into account the contribution of the relativistic
Wigner rotation matrices of the quark spin.
\par
The first term in~(\ref{three_part_mag_mom}) has the form:
\be
 \mu_1 =
 3 \,\, ( \chi \hat{e}_3 \sigma^{(3)}_3 \chi)
 \int d\Omega_{\p_1} d\Omega_{\p_2}|\varphi(\p_1,\p_2,\p_3)|^2
  (E_{\p_1} + E_{\p_2} + E_{\p_3})
  /E^2_{\p_3}.
\label{first_part_mu}
\ee
Charge coefficients
$(\eta^{MS}, \hat{e}_3 \eta^{MS})$ and
$(\eta^{MA}, \hat{e}_3 \eta^{MA})$
are equal to $0$ and $2/3$, respectively, for proton
and $1/3$ and $-1/3$ for neutron;
spin coefficients are as follows:
$(\xi^{MA} \sigma^{(3)}_3 \xi^{MA}) = 1 $,
$(\xi^{MS} \sigma^{(3)}_3 \xi^{MS}) = -1/3 $.
Isospin-spin coefficients in~(\ref{first_part_mu}) are:
$(\chi \hat{e}_3\sigma^{(3)}_3 \chi)^p = 1/3$,
$(\chi\hat{e}_3\sigma^{(3)}_3 \chi)^n =-2/9.$
\par
Using~(\ref{condition_varphi_delta}),
from~(\ref{q_curr_apprx}) and~(\ref{mag_mom_after_condition})
for $\mu_2$ we have:
\be
 \mu_2 =
 - ( \chi \hat{e}_3 \sigma^{(3)}_3 \chi )
 \int d\Omega_{\p_1} d\Omega_{\p_2}
 |\varphi(\p_1,\p_2,\p_3)|^2
  (E_{\p_1} + E_{\p_2} + E_{\p_3})
 (E_{\p_3} - m )/E^3_{\p_3}.
\label{second_part_mu}
\ee
\par
The term $\mu_3$ is represented as a sum of two parts
$\mu_3 = \mu_3^{(1)} + \mu_3^{(2)}$,
where $\mu_3^{(1)}$ originates from differentiation
of Wigner rotation matrices
$\D(L^{-1}_\bK,\p_k)$ ($k=1,2,3$),
$\mu_3^{(2)}$ originates from differentiation
of Wigner rotation matrices
$
\Dp(L_{\po_{12}},\ponot_k)
$ ($k=1,2$).
In these terms differentiation of factors, which
do not contain Pauli matrices, does not contribute
in the nucleon magnetic moments.
Indeed, we perform differentiation under the first component of $\bK$
momentum, but the third term in~(\ref{q_curr_apprx}) contains
the second component of $\p_3$-momentum.
This leads to expressions, which are proportional to
$\delta_{12}=0$, if to apply~(\ref{condition_varphi_delta}).
Taking into account that $\ponot_1=-\ponot_2$, it it possible 
to show that
$\mu_3^{(2)}=0.$ So, we have $\mu_3 = \mu_3^{(1)}.$
\be
 \mu_3 =
  - \int d\Omega_{\p_1} d\Omega_{\p_2}|\varphi(\p_1,\p_2,\p_3)|^2
   \frac{1}{E^2_{\p_3}}
\sum_{k=1}^{3}
 \left[
(\chi \hat{e}_3 \sigma^{(k)}_3 \chi)
  \frac{(\p_k\p_3)}{E_{\p_k} + m}
 \right].
 \label{mu_3}
\ee
For proton and neutron we have used charge coefficients listed above.
Spin coefficients
$(\xi^{MA} \sigma^{(k)}_3 \xi^{MA})$
for $k=1$, $2$ and  $3$ are equal to
$0$, $0$ and $1$, respectively, and
$(\xi^{MS} \sigma^{(k)}_3 \xi^{MS})$
for $k=1$, $2$ and  $3$ are equal to
$2/3$, $2/3$ and $-1/3$,
So, in~(\ref{mu_3}) isospin-spin coefficients
$(\chi\hat{e}_3\sigma^{(k)}_3\chi)$ are equal to
$0$, $0$ and $1/3$ for proton and
$1/9$, $1/9$ and $-2/9$ for neutron.
\par
Using decomposition of the intergrand function in neighborhood
of the effective mass and permutation symmetry
of~(\ref{varphi_approx_eff_m})
in expressions~(\ref{first_part_mu})-~(\ref{mu_3}),
it is possible to change $E_{\p_k}$ to $E_{\p_3}$.
Taking into account the wave normalization
condition~(\ref{norm_varphi}), we obtain a simple result, which
approximates the expressions~(\ref{three_part_mag_mom})-~(\ref{mu_3}):
\be
\mu^p \approx 3-4a/3,\quad
\mu^n \approx -2+a;
\label{mu_approx}
\ee
\be
\mu_1^p \approx 3,\;
\mu_1^n \approx -2;\quad
\mu_2^p \approx -a,\;
\mu_2^n \approx -2a/3;\quad
-\mu_3^p \approx \mu_3^n \approx a/3,
\label{mu123_approx}
\ee
\be
 a=\int d\Omega_{\p_1} d\Omega_{\p_2}|\varphi(\p_1,\p_2,\p_3)|^2
    (E_{\p_3} - m)/E^2_{\p_3}
\label{a_approx}
\ee
In the oscillator-like model~(\ref{varphi_approx_eff_m}) 
with $\gamma/m=0.55,$
corresponding to the best fit of the nucleon magnetic moments,
the deviation of the approximate value from
the exact value is less than $0.1\%$ for $\mu_1$,
less than $0.08\%$ for $\mu_2$ and
less than $0.003\%$ for $\mu_3$
(percent is relative $\mu_N$).
\par
Using the experimental data for the nucleon magnetic moments,
we obtain
$ a_p = 0.155\, (\gamma/m = 0.64)$,
$a_n = 0.087\, (\gamma/m = 0.45)$.
In brackets we have listed corresponding parameters of 
the oscillator-like
model~(\ref{varphi_approx_eff_m}).
For average $a = (a_p +a_n)/2 = 0.121 \, (\gamma/m = 0.55)$
we have:
\ben
  \mu^p &=& 2.854, 2.839,\, ( 2.793),
\nonumber \\
  \mu^n &=&-1.890,-1.879,\, (-1.913).
\label{mu_p_n}
\een
The first two values in the strings are exact
(expressions~(\ref{three_part_mag_mom})-~(\ref{mu_3}))
and approximate values (~(\ref{mu_approx}) for $a=0.121$),
the experimental data are given in brackets.

\begin{center}
{\bf Conclusion.}
\end{center}
\par
The nucleon electromagnetic current in the quasipotential model
with the fixed number of particles is approximately conserved 
in the physically interesting region of the oscillator parameters.
\par
The main result of the work is contained in formulae for the
nucleon magnetic moments:
exact~(\ref{three_part_mag_mom})-~(\ref{mu_3})
and approximate expressions~(\ref{mu_approx})-~(\ref{a_approx}).
The expressions~(\ref{mu_approx}) show, that in the constructed
quasipotential model the main contibution in the nucleon magnetic moment
is defined by the $SU(6)$ part of the nucleon wave function.
The corrections have the right sign for proton and neutron, respectively.
In the constructed model with point quarks without anomalous magnetic
moments the nucleon magnetic moments are consistent with
the experimental data with the error of 2\%.
\par
The independence of the nucleon magnetic moments on the
nucleon mass means that using the experimental data it is possible to
determine only ratio  $\gamma/m$ (for the oscillator-like model).
As a result the mass of the constituent quark is a free parameter.
This conclusion is a consequence of relativistic kinematics and
is not related with the property of the quasipotential model.
\par
Authors gratefully acknoledge V.A.Petrun'kin
for fruitful discussions.

\end{document}